\documentclass{eptcs}

\providecommand{\event}{FESCA 2017}

\newlength{\smallskipamounttt}
\newlength{\organisationfigure}
\usepackage[T1]{fontenc}
\usepackage[latin1]{inputenc}
\usepackage[english]{babel}
\usepackage{%
 amssymb,delarray,graphicx,hufflen-macros,ifpdf,rotating,url,wrapfig}

\newcommand{\Dash}{---}
\newcommand{\ddef}{Def.}
\newcommand{\eex}{Ex.}
\newcommand{\ffig}{Fig.}
\newcommand{\ffigs}{Figs.}
\newcommand{\jmhlineonetitle}{Checking Properties along Multiple}
\newcommand{\jmhlinetwotitle}{Reconfiguration Paths for Component-Based
Systems}
\newcommand{\jmhtitle}{\jmhlineonetitle \\ \jmhlinetwotitle}
\newcommand{\remk}{Rem.}
\newcommand{\surra}[1]{\langle\mathit{#1}\rangle}
\newcommand{\uptoi}[2]{#1_{#2}^{\uparrow}}

\newtheorem{definition}{Definition}
\newtheorem{example}[definition]{Example}
\newtheorem{proposition}[definition]{Proposition}
\newtheorem{remark}[definition]{Remark}

\title{\jmhtitle}
\author{Jean-Michel Hufflen \institute{\logo{femto-st} (\logo{umr}
\logo{cnrs}~6174) \&\ University of Burgundy Franche-Comté \\ 16, route de
Gray; 25030 Besançon Cedex; France} \email{jmhuffle@femto-st.fr}}

\def\titlerunning{\jmhlineonetitle\ \jmhlinetwotitle}
\def\authorrunning{J.-M. Hufflen}

\begin{document}

\maketitle

\begin{abstract}
Reconfiguration paths are used to express sequences of successive
reconfiguration operations within a component-based approach allowing dynamic
reconfigurations. We use constructs from regular expressions\Dash in
particular, alternatives\Dash to introduce \emph{multiple reconfiguration
paths}. We show how to put into action procedures allowing architectural,
event, and temporal properties to be proved. Our method, related to finite
state automata and using marking techniques, generalises what we did within
previous work, where the regular expressions we processed were more restricted.
But we can only deal with a subset of first-order logic formulas.

\end{abstract}

\section{Introduction}

Dynamic reconfigurations of software architectures are active research topics
\cite{allen-etc1998,bozga-etc2012,bruni-lavanese2006,krause-etc2011,lanoix-kouchnarenko2014,leger-etc2010,kouchnarenko-weber2014,sanchez-etc2015}.
They provide large increase in value for component-based software. Such an
approach allows some components to be replaced or removed, in particular if
they fail. In order to provide more services, more components may be added
dynamically, too. So dynamic reconfigurations increase the availability and
reliability of such systems by allowing their architecture to evolve at
run-time. 

The work presented hereafter is an extension of \cite{h2015a}, which addresses
the verification of \emph{architectural}, \emph{event}, or \emph{temporal}
properties. Such properties may be crucial for systems with high-safety
requirements. About the definition of such properties, \cite{dormoy-etc2010}
proposes \logo{ftpl}\footnote{\textbf{F}ractal \textbf{T}emporal
\textbf{P}attern \textbf{L}ogic.}, a temporal logic for dynamic
reconfigurations applied to components defined by means of the \pgname{Fractal}
toolbox \cite{bruneton-etc2006} and including such properties. \logo{ftpl}
allows successive reconfigurations\Dash modelled by \emph{reconfiguration
paths}\Dash to be applied to successive \emph{configurations} (or
\emph{component models}). Since \logo{ftpl} is based on first-order predicate
logic, such properties are undecidable in general, there only exist partial
solutions for proving them.

Many authors developed methods that work whilst software is running and may be
reconfigured\Dash e.g., \cite{kouchnarenko-weber2013,kouchnarenko-weber2014},
based on \logo{ftpl}, or \cite{falcone-etc2011} as another example. Therefore
we know if a property holds for the successive members of a chain of
reconfigurations, until the current run-time state. Our method is very
different, more related to the approach of a procedure's developer when such a
developer aims to prove its procedure before deploying it and putting it into
action. In fact, we do not verify such properties at run-time, but on a static
abstraction of the reconfiguration model, so we aim to ensure that such a
property holds before the software is deployed and working, that is, at
design-time. Of course, we cannot consider reconfigurations caused by totally
unexpected events but we think that our approach is \emph{complementary} to
such works, our goal is to go as far as possible within this static approach.
In \cite{h2015a}, we proposed a method based on this point of view and using
marking techniques related to model-checking: given a reconfiguration path that
may be applied when the software is running, we aimed to ensure that a property
holds if this path is actually applied when the software works. We were able to
deal with some cases of infinite reconfiguration paths, but we only processed
\emph{one} possible reconfiguration path. Dealing with only one path is not
restrictive for methods applied at run-time, whilst the software is working,
but is rather limited at design-time, where \emph{several} possible futures
could be studied. In the present article, we propose the new notion of
\emph{multiple reconfiguration paths}, which are expressions denoting
\emph{several possible} reconfiguration processes. However, this extension has
a price: the correctness of our new implementations\Dash w.r.t.\ the
definitions of \cite{dormoy-etc2010}\Dash is guaranteed only for a strict
subset of formulas, in comparison with formulas used within \cite{h2015a}.

Section~\ref{dkl-reusing} gives some recalls about the component model we use,
our operations of reconfiguration, and the temporal logic for dynamic
reconfigurations. Of course, most definitions presented in this section come
from
\cite{dormoy-etc2010,dormoy-etc2011,dormoy-etc2012,kouchnarenko-weber2013}.
Section~\ref{multiple-reconfiguration-paths} precisely introduces our notion of
multiple reconfiguration path and Section~\ref{framework} recalls the
organisation of our framework. Then we give updated versions of our programs in
Section~\ref{implementation} and study the correctness of these implementations
w.r.t.\ the operators defined in Section~\ref{dkl-reusing}. We do not examine
all the operators, but our examples are representative: implementation
techniques and correctness proofs are analogous. Section~\ref{discussion}
discusses some advantages and drawbacks of our method, in comparison with other
approaches. It also introduces future work. In order for this article to be
self-contained, most of the definitions put hereafter are identical to
\cite{h2015a}'s. Readers familiar with that article can skip
Section~\ref{dkl-reusing}\Dash except for the definition of the
$\mathit{CP}^{\flat}$ set\Dash and \S~\ref{modus-operandi}.

\section{Architectural Reconfiguration Model} \label{dkl-reusing}

First we recall how our \emph{component model} is organised. Then we sum up the
operations used for reconfiguring an architecture. Last, we make precise
operators used in \logo{ftpl}, the temporal logic used in
\cite{dormoy-etc2010,dormoy-etc2011,dormoy-etc2012,kouchnarenko-weber2013} for
dynamic reconfigurations.

\subsection{Component Model} \label{component-model}

Roughly speaking, a component model describes an \emph{architecture} of
components. Some simpler components may be subcomponents of a \emph{composite}
one, and components may be \emph{linked}. Let $\mathcal{S}$ be a set of
\emph{type names} \footnote{\ldots\ or \emph{class names} within an
object-oriented approach.}, a \emph{component} $\mathcal{C}$ is defined by:
\begin{itemize}
 \item three pairwise-disjoint sets of \emph{parameters}\footnote{Some authors
use the term `attributes' instead. A parameter is related to an internal
feature, e.g., the maximum number of messages a component can process.}
$P_{\mathcal{C}}$, \emph{input port} names $I_{\mathcal{C}}$, and \emph{output
port} names $O_{\mathcal{C}}$;
 \item the class $t_{\mathcal{C}}$ encompassing the services implemented by the
component;
 \item additional functions to get access to the class of a parameter or port
($\tau_{C} : P_{\mathcal{C}} \cup I_{\mathcal{C}} \cup O_{\mathcal{C}}
\rightarrow \mathcal{S}$), or to a parameter's value ($v_{\mathcal{C}} :
P_{\mathcal{C}} \rightarrow \bigcup_{s \in \mathcal{S}} s$);
 \item the set $\mathit{sub\text{-}c}_{\mathcal{C}}$ of its subcomponents if 
the $\mathcal{C}$ component is composite\footnote{Of course, the binary
relation `is a subcomponent of' must be a direct acyclic graph. A composite
component cannot have parameters. More precisely, it implicitly has the
parameters of all its sub-components.};
 \item the set $B$ of \emph{bindings} of ports\Dash that is, couples of input
and input port names, being the same type, and the set $D$ of \emph{delegation
links}, between composite component ports and port of contained components.
\end{itemize}

\begin{figure}[t]

\begin{center}
\ifpdf\includegraphics[width=0.9\linewidth]{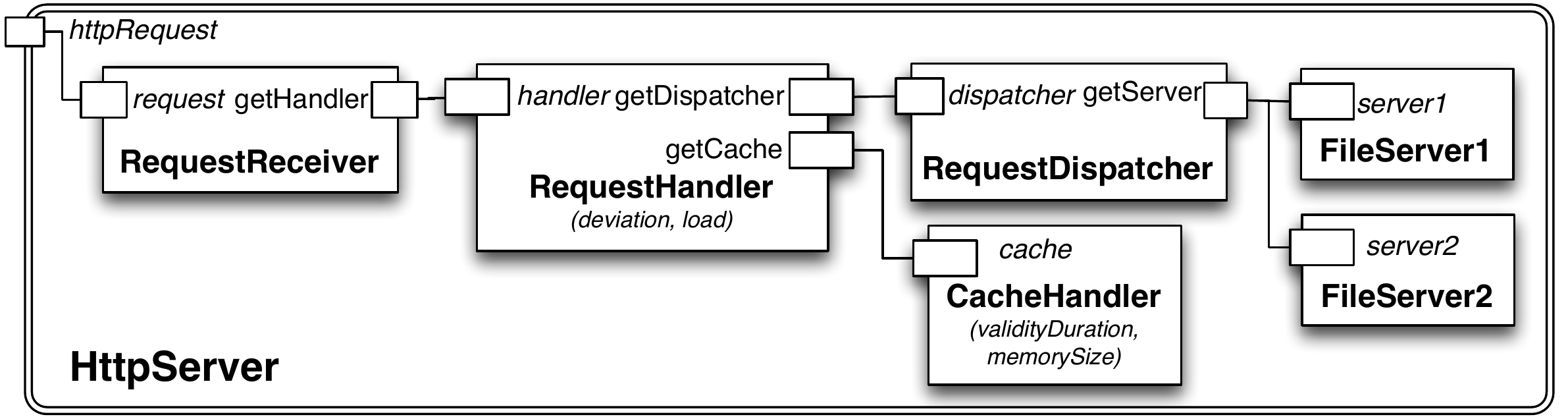}\fi
\end{center}

\caption{Component-based architecture of an \logo{http} server
\cite{dormoy-etc2012}.} \label{hstr-http-server}
\frule
\end{figure}

Possible components of an \logo{http} server are given in
\ffig~\ref{hstr-http-server}, as an example of a component-based architecture,
already used in \cite{chauvel2008a}. Requests are read by the
\texttt{RequestReceiver} component and transmitted to the
\texttt{RequestHandler} component. When the latter processes a request, it may
consult the cache by means of the \texttt{CacheHandler} component or transmit
this request to the \texttt{RequestDispatcher} component, which manages file
servers. This architecture is based on a cache and load balancer, in order for
response times to be as short as possible. The cache must be used only if the
number of similar requests is very high, and the amount of memory devoted to
the cache component must be automatically adjusted to the Web server's load.
The validity duration of the data put in the cache must also be adjusted with
respect to the Web server's load. In addition, more data servers have to be
deployed if the servers' average load is high. According to these conventions,
we see that some components may be added or removed, depending on some
parameters.

\subsection{Configuration Properties} \label{cp-flat}

\begin{example} \label{cache-connected}
Looking at \ffig~\ref{hstr-http-server}'s architecture, we can notice that the
\emph{\texttt{CacheHandler}} component is connected to the
\emph{\texttt{RequestHandler}} component through their respective ports
\emph{\texttt{cache}} and \emph{\texttt{getCache}}. We can express this
\emph{configuration property}\Dash so-called
\emph{\texttt{CacheConnected}}\Dash as follows:
\[B \ni
(\text{\emph{\texttt{cache}}}_{\text{\emph{\texttt{CacheHandler}}}},\text{\emph{\texttt{getCache}}}_{\text{\emph{\texttt{RequestHandler}}}})\]
\end{example}

In fact, such properties\Dash that may be viewed as \emph{constraints}\Dash are
specified using first-order logic formulas over constants (`\texttt{true}',
`\texttt{false}'), variables, sets and functions defined
in~\S~\ref{component-model}, predicates ($=,\in,\ldots$), connectors
($\wedge,\vee,\ldots)$ and quantifiers ($\forall,\exists$). These configuration
properties form a set denoted by $\mathit{CP}$. The subset
$\mathit{CP}^{\flat}$ is build analogously, but connectors and quantifiers are
restricted to $\wedge$ and $\forall$. Roughly speaking, formulas belonging to
$\mathit{CP}^{\flat}$ are comparable to premises of Horn clauses within logic
programming.

\subsection{Reconfiguration Operations} \label{reconfiguration}

\emph{Primitive} reconfiguration operations apply to a component architecture,
and the output is a component architecture, too\footnote{They may be viewed as
\emph{graph transformations} applied to component models if we consider such
models as graphs.}. They are the addition or removal of a component, the
addition or removal of a binding, the update of a parameter's value. Let us
notice that the result of such an operation is consistent from a point of view
related to software architecture: for example, a component is stopped before it
is removed, and removing it causes all of its bindings to be removed, too.
These operations are robust in the sense that they behave like the identity
function if the corresponding operation cannot be performed. For example, if
you try to remove a component not included in an architecture, the original
architecture will be returned. The same if you try to add a component already
included in the architecture\footnote{The reason: the \emph{name} of a
component\Dash part of its definition\Dash can only identify \emph{one}
component. But you can \emph{clone} a component under a new name.}. As a
consequence, these \emph{topological} operations\Dash addition or removal of a
component or a binding\Dash are \emph{idempotent}: applying such an operation
twice results in the same effect than applying it once. General reconfiguration
operations on an architecture are combinations of primitive ones, and form a
set denoted by $\mathcal{R}$. The set of \emph{evolution operations} is
$\mathcal{R}_{\mathit{run}} = \mathcal{R} \cup \{\mathit{run}\}$ where
$\mathit{run}$ is an action modelling that all the stopped components are
restarted and the software is running\footnote{Strictly speaking, we have to
\emph{stop} a component before removing it, and to \emph{start} it before
having added it, as abovementioned. This convention about the $\mathit{run}$
action allows us not to be worried about such \emph{stop} and \emph{start}
operations within our reconfiguration paths.}.

\begin{definition}[\cite{dormoy-etc2011,kouchnarenko-weber2013}]
\label{operational-semantics}
The operational semantics of component systems with reconfigurations is defined
by the labelled transition system $\mathcal{S} =
\surra{C,C^{0},\mathcal{R}_{\mathit{run}},\rightarrow,l}$ where $C =
\{c,c_{1},c_{2},\ldots\}$ is a set of configurations\Dash or component
models\Dash $C^{0} \subseteq C$ is a set of initial configurations,
$\mathcal{R}_{\mathit{run}}$ is a finite set of evolution operations,
$\rightarrow\; \subseteq C \times \mathcal{R}_{\mathit{run}} \times C$ is the
reconfiguration relation, and $l : C \rightarrow \mathit{CP}$ is a total
function to label each $c \in C$ with the largest conjunction of $\mathit{cp}
\in \mathit{CP}$ evaluated to `true' over $\mathcal{R}_{\mathit{run}}$.
\end{definition}

\begin{figure}[t]

\begin{center}
\ifpdf\includegraphics[width=0.9\linewidth]{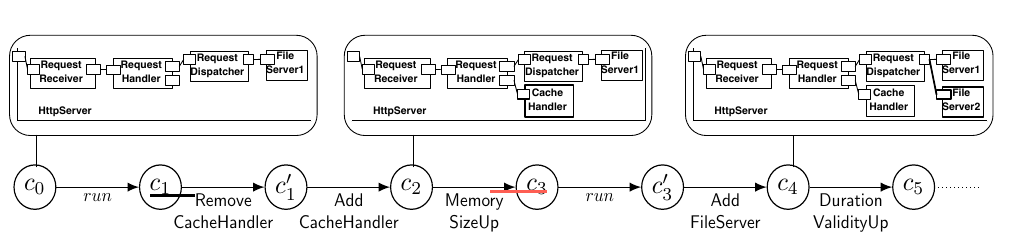}\fi
\end{center}

\caption{Part of an evolution path of \ffig~\protect\ref{hstr-http-server}'s
\logo{http} server architecture \cite{dormoy-etc2012}.} \label{hstr-http-path}
\frule
\end{figure}

Let us note $c \stackrel{\mathit{op}}{\rightarrow} c'$ when a target
configuration $c'$ is reached from a configuration $c$ by an evolution
$\mathit{op} \in \mathcal{R}_{\mathit{run}}$. Given the model $S =
\surra{C,C^{0},\mathcal{R}_{\mathit{run}},\rightarrow,l}$, an evolution path
$\sigma$ of $S$ is a (possibly infinite) sequence of component models
$c_{0},c_{1},c_{2},\ldots$ such that $\forall i \in \mathbb{N}, \exists
\mathit{op} \in \mathcal{R}_{\mathit{run}}, c_{i}
\stackrel{\mathit{op}}{\rightarrow} c_{i + 1} \in\; \rightarrow$. We write
`$\sigma[i]$' to denote the $i$th element of a path $\sigma$, if this element
exists. The notation `$\uptoi{\sigma}{i}$' denotes the suffix path
$\sigma[i],\sigma[i + 1],\ldots$ and `$\sigma_{i}^{j}$' ($j \in \mathbb{N}$)
denotes the segment path $\sigma[i],\sigma[i + 1],\ldots,\sigma[j -
1],\sigma[j]$. An example of evolution path allowing
\ffig~\ref{hstr-http-server} to be reached from a simpler architecture is given
in \ffig~\ref{hstr-http-path} (\ffig~\ref{hstr-http-server}'s architecture is
labelled by the $c_{4}$ configuration).

\subsection{Temporal Logic} \label{temporal-logic}

\logo{ftpl} deals with events from reconfiguration operations, trace
properties, and temporal properties, respectively denoted by
`$\mathit{event}$', `$\mathit{trace}$', and `$\mathit{temp}$' in the following.
Hereafter we only give some operators of \logo{ftpl}, in particular those used
in the implementations we describe. For more details about this temporal logic,
see \cite{dormoy-etc2011,kouchnarenko-weber2013}. \logo{ftpl}'s syntax is
defined by:
\begin{eqnarray*}
\surra{temp} & ::= & \text{\textbf{after}}\; \surra{event}\; \surra{temp} \mid
\text{\textbf{before}}\; \surra{event}\; \surra{trace} \mid \ldots \\
\surra{trace} & ::= & \text{\textbf{always}}\; \mathit{cp} \mid
\text{\textbf{eventually}}\; \mathit{cp} \mid \ldots \\
\surra{event} & ::= & \mathit{op}\; \text{\textbf{normal}} \mid \mathit{op}\;
\text{\textbf{exceptional}} \mid \mathit{op}\; \text{\textbf{terminates}}
\end{eqnarray*}
where `$\mathit{cp}$' is a configuration property and `$\mathit{op}$' a
reconfiguration operation.
Let $\mathit{cp}$ in $\mathit{CP}$ be a configuration property and $c$ a
configuration, $c$ satisfies $\mathit{cp}$, written `$c \models \mathit{cp}$'
when $l(c) \Rightarrow \mathit{cp}$. Otherwise, we write `$c \not\models
\mathit{cp}$' when $c$ does not satisfy $\mathit{cp}$.

\begin{definition}[\cite{dormoy-etc2011}] \label{evolution-path-df}
Let $\sigma$ be an evolution path, the \logo{ftpl} semantics is defined by
induction on the form of the formulas as follows\footnote{For a complete
definition including all the operators, see \cite{dormoy-etc2011}.}\Dash in the
following, $i \in \mathbb{N}$\Dash:
\end{definition}
\begin{itemize}
 \item for the events:
\begin{center}
\begin{tabular}{l@{\quad if\quad}l}
$\sigma[i] \models \mathit{op}\; \text{\textbf{normal}}$ & $i > 0 \wedge
\sigma[i - 1] \ne \sigma[i] \wedge \sigma[i - 1]
\stackrel{\mathit{op}}{\rightarrow} \sigma[i] \in\; \rightarrow$ \\
$\sigma[i] \models \mathit{op}\; \text{\textbf{exceptional}}$ & $i > 0 \wedge
\sigma[i - 1] = \sigma[i] \wedge \sigma[i - 1]
\stackrel{\mathit{op}}{\rightarrow} \sigma[i] \in\; \rightarrow$ \\
$\sigma[i] \models \mathit{op}\; \text{\textbf{terminates}}$ & $\sigma[i]
\models \mathit{op}\; \text{\textbf{normal}} \vee \sigma[i] \models
\mathit{op}\; \text{\textbf{exceptional}}$
\end{tabular}
\end{center}
 \item for the trace properties:
\begin{center}
\begin{tabular}{l@{\quad if\quad}l}
$\sigma \models \text{\textbf{always}}\; \mathit{cp}$ & $\forall i : i \ge 0
\Rightarrow \sigma[i] \models \mathit{cp}$ \\
$\sigma \models \text{\textbf{eventually}}\; \mathit{cp}$ & $\exists i : i \ge
0 \Rightarrow \sigma[i] \models \mathit{cp}$
\end{tabular}
\end{center}
 \item for the temporal properties:
\begin{center}
\begin{tabular}{l@{\quad if\quad}l}
$\sigma \models \text{\textbf{after}}\; \mathit{event}\; \mathit{temp}$ &
$\forall i : i \ge 0 \wedge \sigma[i] \models \mathit{event} \Rightarrow
\uptoi{\sigma}{i} \models \mathit{temp}$ \\
$\sigma \models \text{\textbf{before}}\; \mathit{event}\; \mathit{trace}$ &
$\forall i : i > 0 \wedge \sigma[i] \models \mathit{event} \Rightarrow
\sigma_{0}^{i - 1} \models \mathit{trace}$
\end{tabular}
\end{center}
\end{itemize}

\begin{example} \label{after-always-example}
If we consider the evolution path of \ffig~\ref{hstr-http-path} again, we can
now express that after calling the \emph{\textsf{AddCacheHandler}}
reconfiguration operation, the \emph{\texttt{CacheHandler}} component is always
connected to the \emph{\texttt{RequestHandler}} component\Dash
\emph{\texttt{CacheConnected}} is the configuration property defined in
Example~\ref{cache-connected}\Dash:
\begin{center}
\emph{\textbf{after \textnormal{\textsf{AddCacheHandler}} normal always
\textnormal{\texttt{CacheConnected}}}}
\end{center}
\end{example}

\begin{remark}
About temporal and trace properties, let us notice that if such a property
holds on an evolution path, it holds on any prefix of this path.
\end{remark}

\section{Multiple Reconfiguration Paths} \label{multiple-reconfiguration-paths}

\begin{definition} \label{multiple-reconfiguration-path-df}
Let $\mathcal{R}_{\mathit{run}}$ be a set of \emph{evolution operations}, a
\emph{\textbf{reconfiguration path}} is a sequence of elements of
$\mathcal{R}_{\mathit{run}}$, and the set $\Omega_{\mathcal{R}_{\mathit{run}}}$
of \emph{\textbf{multiple}} reconfiguration paths on
$\mathcal{R}_{\mathit{run}}$ is the set of \emph{regular expressions} built
over the alphabet $\mathcal{R}_{\mathit{run}}$. Let us recall that the
constructs used within regular expressions are `\emph{\texttt{\makebar}}' for
alternatives, `\emph{\texttt{?}}' for an optional occurrence of an alphabet's
member, `\emph{\texttt{*}}' (resp.\ `\emph{\texttt{+}}') for zero (resp.\ one)
or more occurrences of such a member. Semantically, a multiple reconfiguration
path is the set of all the prefixes of all the reconfiguration paths denoted by
this regular expression.
\end{definition}

\begin{example} \label{multiple-reconfiguration-path-ex}
The following multiple reconfiguration path:
\begin{ttfamily}
\begin{tabbing}
$\mathit{run}$ \textnormal{\textsf{RemoveCacheHandler AddCacheHandler}} \\
(\= \textnormal{\textsf{MemorySizeUp}} $\mathit{run}$ \\
 \> (\textnormal{\textsf{AddFileServer DurationValidityUp}} \emph{\makebar}
\textnormal{\textsf{DurationValidityUp AddFileServer}}) $\mathit{run}$?\ \\
 \> \textnormal{\textsf{DeleteFileServer}})+
\textnormal{\textsf{AddFileServer}}
\end{tabbing}
\end{ttfamily}
includes the chain of reconfigurations pictured at \ffig~\ref{hstr-http-path}.
\end{example}

\begin{remark}
Let us recall that a reconfiguration path may be infinite. Looking at
\eex~\ref{multiple-reconfiguration-path-ex}, we consider that the
`\emph{\texttt{(\ldots)+}}' expression can be iterated a finite number of
times, followed by the \emph{\textsf{AddFileServer}} operation; another
possible behaviour is an endless iteration of the `\emph{\texttt{(\ldots)+}}'
expression. We encompass all these possible behaviours by considering prefixes,
as mentioned in \ddef~\ref{multiple-reconfiguration-path-df}.
\end{remark}

\begin{figure}[t]

\begin{center}
\begin{picture}(30,14)

\put(1,6){\circle{2}}
\put(1,2){\circle{2}}
\put(4.5,2){\circle{2}}
\put(8,2){\circle{2}}
\put(11.5,2){\circle{2}}
\put(15,2){\circle{2}}
\put(18.5,2){\circle{2}}
\put(18.5,6){\circle{2}}
\put(22,2){\circle{2}}
\put(25.5,2){\circle{2}}
\put(29,2){\circle{2}}

\put(1,6){\makebox(0,0){$q_{0}$}}
\put(1,2){\makebox(0,0){$q_{1}$}}
\put(4.5,2){\makebox(0,0){$q'_{1}$}}
\put(8,2){\makebox(0,0){$q_{2}$}}
\put(11.5,2){\makebox(0,0){$q_{3}$}}
\put(15,2){\makebox(0,0){$q'_{3}$}}
\put(18.5,2){\makebox(0,0){$q_{4}$}}
\put(18.5,6){\makebox(0,0){$q_{7}$}}
\put(22,2){\makebox(0,0){$q_{5}$}}
\put(25.5,2){\makebox(0,0){$q_{6}$}}
\put(29,2){\makebox(0,0){$q_{8}$}}

\put(1,5){\vector(0,-1){2}}
\put(2,2){\vector(1,0){1.5}}
\put(5.5,2){\vector(1,0){1.5}}
\put(9,2){\vector(1,0){1.5}}
\put(12.5,2){\vector(1,0){1.5}}
\put(16,2){\vector(1,0){1.5}}
\put(15.4,3){\vector(2,3){2}}
\put(19.5,2){\vector(1,0){1.5}}
\put(19.5,6){\vector(2,-3){2}}
\put(23,2){\vector(1,0){1.5}}
\put(26.5,2){\vector(1,0){1.5}}
\put(15,1){\oval(14,1)[b]}
\put(16.75,1){\oval(17.5,2)[b]}
\put(8,0.9){\vector(0,1){0.1}}

\put(0.8,4){\makebox(0,0)[r]{\begin{turn}{90}
$\mathit{run}$
\end{turn}}}
\put(2.8,2.5){\makebox(0,0)[b]{\begin{turn}{90}
\textsf{RemoveCacheHandler}
\end{turn}}}
\put(6.3,2.5){\makebox(0,0)[b]{\begin{turn}{90}
\textsf{AddCacheHandler}
\end{turn}}}
\put(9.8,2.5){\makebox(0,0)[b]{\begin{turn}{90}
\textsf{MemorySizeUp}
\end{turn}}}
\put(13.3,2.5){\makebox(0,0)[b]{\begin{turn}{90}
$\mathit{run}$
\end{turn}}}
\put(16.8,2.5){\makebox(0,0)[b]{\begin{turn}{90}
\textsf{D\ldots}
\end{turn}}}
\put(16.8,5.8){\makebox(0,0)[b]{\begin{turn}{90}
\textsf{AddFileServer}
\end{turn}}}
\put(20.3,2.5){\makebox(0,0)[b]{\begin{turn}{90}
\textsf{A\ldots}
\end{turn}}}
\put(20.3,6){\makebox(0,0)[b]{\begin{turn}{90}
\textsf{DurationValidityUp}
\end{turn}}}
\put(23.8,2.5){\makebox(0,0)[b]{\begin{turn}{90}
$\mathit{run}$
\end{turn}}}
\put(27.3,2.5){\makebox(0,0)[b]{\begin{turn}{90}
\textsf{AddFileServer}
\end{turn}}}
\put(16.75,0){\makebox(0,0)[t]{\textsf{DeleteFileServer}}}

\end{picture}

\end{center}

\caption{Automaton for a multiple reconfiguration path.} \label{hstr-automaton}
\frule
\end{figure}
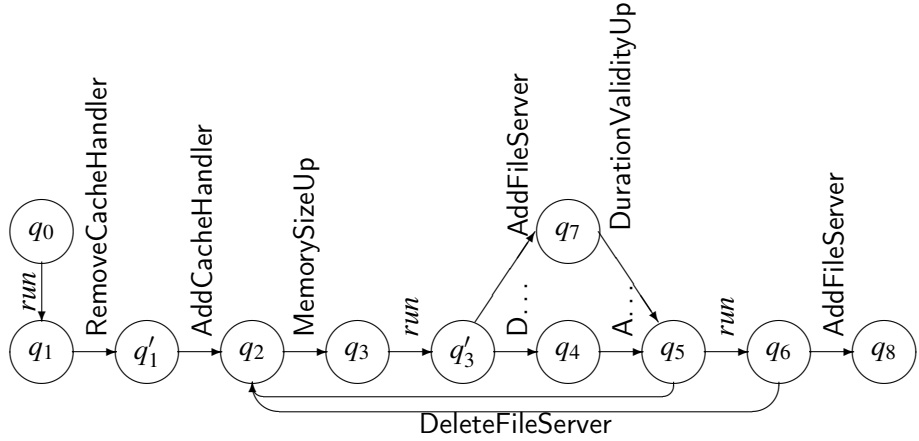

It is well-known for many years\Dash since Kleene's theorem\Dash that a regular
expression language can be recognised by a deterministic finite state
automaton, whose transitions are labelled by members of this language's
alphabet. Let us recall that such an automaton $\mathcal{A}$ is defined by a
set $Q$ of \emph{states}, a set $L$ of \emph{transition labels}, and a set $T
\subseteq Q \times L \times Q$ of \emph{transitions}. As in
\ddef~\ref{evolution-path-df} for systems with reconfigurations, there exists a
function $l : Q \rightarrow \mathit{CP}$, which labels each $q$ state with the
largest conjunction of $\mathit{cp} \in \mathit{CP}$ evaluated to `true' for
the $q$ state. As an example, \eex~\ref{multiple-reconfiguration-path-ex}'s
language can be recognised by the automaton pictured in
\ffig~\ref{hstr-automaton} (the states $q_{0},q_{1},q'_{1},\ldots,q_{5}$ have
been respectively named in connection to the successive component models
$c_{0},c_{1},c'_{1},\ldots,c_{5}$ of \ffig~\ref{hstr-http-path}). In addition,
let us recall that such an automaton can be build automatically from a regular
expression. In the next section, we explain what our states are, and which
operations are performed by our transitions.

\section{Our Method's Bases} \label{framework}

\subsection{\emph{Modus Operandi}} \label{modus-operandi}

As mentioned above, our framework's basis is an automaton modelling the
possible evolution paths of a multiple reconfiguration path. A state of such an
automaton is a component model, initial or got by means of successive
reconfiguration operations\Dash primitive or built by chaining primitive
operations\Dash or `\emph{run}' operations. A transition consists of applying
such an evolution operation. Such an automaton has an initial state, given by
the initial component model ($q_{0}$ in \ffig~\ref{hstr-automaton}). Since we
aim to recognise all the prefixes of possible reconfiguration paths, any state
may be viewed as final. In addition, since some infinite behaviours are
accepted (e.g., endlessly cycling from the $q_{5}$ or $q_{6}$ state to the
$q_{2}$ state in \ffig~\ref{hstr-automaton}), there are processes without
`actual' final state. In fact, the complete automaton may be viewed as an
$\omega$-automaton. Let us go back to states reached several times\Dash e.g.,
the $q_{2}$ state in \ffig~\ref{hstr-automaton}, reached after $q_{5}$ and
$q_{6}$\Dash: considering that the whole system is back to a previous state may
be not exact, because some parameters may have been updated: this is the case
in \ffig~\ref{hstr-automaton}'s example, about the memory's size and duration
validity. As a consequence, some properties related to components' parameters
may not hold. We will go back on this point at the beginning of
\S~\ref{cp-flat}.

\begin{wrapfigure}[11]{l}{\organisationfigure}

\centering\begin{picture}(15,7.6)

\put(0,1){\framebox(4,2){{\small $c_{p}$}}}
\put(7,1){\framebox(4,2){{\small $c_{p + 1}$}}}
\put(4,2){\vector(1,0){3}}
\put(11,2){\vector(1,0){3}}
\put(2,7.6){\vector(0,-1){4.6}}
\put(9,7.6){\vector(0,-1){4.6}}
\put(2.2,5.3){\makebox(0,0)[l]{{\small \shortstack[l]{program checking \\ a
property, \\ returning \\ \texttt{true} or \\ \texttt{false}}}}}
\put(5.5,1.5){\makebox(0,0)[t]{{\small \shortstack[c]{\logo{xslt} \\
stylesheet}}}}
\put(14.2,2){\makebox(0,0)[l]{\ldots}}

\end{picture}

\caption{Our organisation.} \label{figure-organisation}
\frule[\organisationfigure]
\end{wrapfigure}
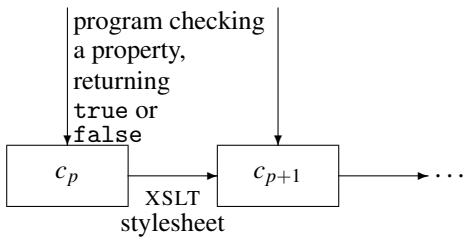

Several programming languages are used within our framework.
\ffig~\ref{figure-organisation} shows how tasks are organised within our
architecture\Dash $(c_{p})_{p \in \mathbb{N}}$ being successive component
models. In our implementation, the \logo{adl}\footnote{\textbf{A}rchitecture
\textbf{D}efinition \textbf{L}anguage.} we use for our component models is
\logo{tacos}+/\logo{xml} \cite{h2013h}. This language using
\logo{xml}\footnote{e\textbf{X}tensible \textbf{M}arkup
\textbf{L}anguage.}-like syntax is comparable with other \logo{adl}s, in
particular \pgname{Fractal}/\logo{adl} \cite{bruneton-etc2006}, but we mention
that the organisation of \logo{tacos}+/\logo{xml} texts make very easy the
programming of primitive reconfiguration operations mentioned
in~\S~\ref{reconfiguration}, that is why we chose this \logo{adl}, a short
example is given in \cite{h2015a}. Reconfigurations operations are implemented
using \logo{xslt}\footnote{e\textbf{X}tensible \textbf{S}tylesheet
\textbf{L}anguage \textbf{T}ransformations, the language of transformations
used for \logo{xml} documents \cite{wwwc2007a}. Let us note that if another
\logo{adl} is used within a project, there exist \logo{xslt} programs giving
equivalent descriptions in \logo{tacos}/\logo{xml} \cite{h2013h}. In
particular, that is the case for \pgname{Fractal}/\logo{adl}.}: the input and
output are \logo{tacos}+/\logo{xml} files.

When we model that the software is running, only one component model is in use,
so that may be viewed as the identity function applied to a component model. In
the programs given below, we compute each component model belonging to a
reconfiguration path. For each component model, we may verify topological
properties, e.g., checking that a component or binding is present. As in
\cite{h2015a}, these topological properties are computed by means of
\pgname{XQuery} programs \cite{wwwc2008d}. There is no difficulty about the
implementation of reconfiguration operations and property checks, so the
descriptions put hereafter concern the part implemented by means of automata.

\subsection{Types Used} \label{types-used}

Now we describe our checking functions at a high level. First we make precise
the types used, in order to ease the reading of our functions. The formalism we
use is close to type definitions in strong typed functional programming
languages like \pgname{Standard ML} \cite{paulson1996} or \pgname{Haskell}
\cite{marlow2010}. Of course, we assume that some types used hereafter\Dash
e.g., `\texttt{bool}', `\texttt{int}'\Dash are predefined. We use the same
names than in \cite{h2015a} for identical notions, and new functions introduced
are suffixed by `\texttt{*}' or `\texttt{**}'.

As mentioned above, an evolution operation is either the identity function,
which expresses that the software is running, or a reconfiguration operation,
which is implemented by applying an \logo{xslt} stylesheet to an \logo{xml}
document and getting the result as another \logo{xml} document. At a
higher-level, such an evolution operation may be viewed as a function which
applies to a component model and returns a component model. Likewise, checking
a property may be viewed as a function which applies to a component model and
returns a boolean value. Assuming that the \texttt{component-model} type has
already been defined, we introduce these two function types as:
\begin{ttfamily}
\begin{center}
\begin{tabular}{l@{ = }l}
\textnormal{\textbf{type}} evolution-op & component-model $\rightarrow$
component-model \\
\textnormal{\textbf{type}} check-property & component-model $\rightarrow$ bool
\end{tabular}
\end{center}
\end{ttfamily}
An \textbf{event} is defined by an evolution operation and a symbol related to
this operation's result (\cf~\ddef~\ref{evolution-path-df}):
\begin{ttfamily}
\begin{center}
\begin{tabular}{l@{ }c@{ }l@{ $\rightarrow$ }l}
\textnormal{\textbf{function}} event->ev-op & : & event & evolution-op \\
\textnormal{\textbf{function}} event->termination-s & : & event &
termination-symbol \\
\textnormal{\textbf{type}} termination-symbol & = &
\multicolumn{2}{@{}l}{\{normal,exceptional,terminates\}}
\end{tabular}
\end{center}
\end{ttfamily}
This last information is used by a function checking that the component model
got by an evolution operation and the previous component model are equal or
different, depending on this symbol\footnote{Let us recall
(\cf~\ddef~\ref{evolution-path-df}) that if this symbol is
`\texttt{terminates}', no additional checking is performed.}:
\begin{center}
\textbf{function} \texttt{term-check : event $\rightarrow$ (component-model
$\times$ component-model $\rightarrow$ bool)}
\end{center}
Let \texttt{state} be the type used for a state of our automata, starting from
such a state and a configuration\footnote{That is, a component model (see
\ddef~\ref{operational-semantics}).} is expressed by the following type:
\begin{center}
\texttt{\textnormal{\textbf{type}} path-check = state $\times$ component-model
$\rightarrow$ bool}
\end{center}
The following function yields all the transitions starting from a state:
\begin{center}
\texttt{\textnormal{\textbf{function}} t : state $\rightarrow$
set-of[transition]}
\end{center}
the data belonging to a set can be accessed by means of a `\textbf{for}'
expression. A transition starts from a state and returns a state, and the label
of such a transition is given by the \texttt{l} function:
\begin{center}
\begin{ttfamily}
\begin{tabular}{r@{ $\rightarrow$ }l}
\textnormal{\textbf{type}} transition = state & state \\
\textnormal{\textbf{function}} l : transition & evolution-op
\end{tabular}
\end{ttfamily}
\end{center}
In the following, we will focus on the constructs `\textbf{after}' and
`\textbf{always}'. The \texttt{path-check} type is used within:
\begin{ttfamily}
\begin{center}
\begin{tabular}{l@{ : }r@{ $\rightarrow$ }l}
\textnormal{\textbf{function}} check-after* & evolution-op $\times$ path-check
& path-check \\
\textnormal{\textbf{function}} check-always* & check-property & path-check
\end{tabular}
\end{center}
\end{ttfamily}
In other words, \texttt{check-always*($\mathit{check\text{-}p*}$)($q$,$c$)}
applies the $\mathit{check\text{-}p*}$ function along the $q$ state, the states
reached by transitions originating from $q$, and so on, starting from the $c$
component model. The result of this expression is a boolean value. As soon as
applying the $\mathit{check\text{-}p*}$ function yields `false', the process
stops and the result is `false'. Likewise,
\texttt{check-after*($e$,$\mathit{check\text{-}f*}$)($q$,$c$)} also starts from
the $q$ state and the $c$ component model; it applies the
$\mathit{check\text{-}f*}$ function as soon as the $e$ event is detected as a
transition of the automata. The property related to the
$\mathit{check\text{-}f*}$ function is to be checked for all the component
models resulting from the application of the successive transitions. As a more
complete example, the translation of the formula `\textbf{after} $e$
\textbf{always} $\mathit{cp}$'\Dash where $e$ is an event and $\mathit{cp}$ a
configuration property\Dash is
\texttt{check-after*($e$,check-always*($\mathit{cp}$))}, which is a function
that applies on a path, starting from a state and component model. The process
starts from the initial state of the automaton. Of course, there are similar
declarations for functions such as \texttt{check-before*} and
\texttt{check-eventually*} (\cf~\S~\ref{temporal-logic}). 

\subsection{Ordering States of Automata}

In this section, we introduce some notions related to our automata and used in
the following. The states of our automata modelling multiple reconfiguration
paths can be ordered with respect to the transitions performed before cycling.
Let $\mathcal{A}$ be an automaton, $q_{0}$ its initial state, $L$ its set of
transition labels, and $T$ its set of transitions, if $q$ and $q'$ are two
states of $\mathcal{A}$:
\begin{eqnarray*}
q \mapsto q' & \stackrel{\text{def}}{\Longleftrightarrow} & \exists \tau \in T,
\exists l \in L, \tau = (q,l,q') \quad \text{[By language abuse, we note $q' =
\tau(q)$.]} \\
q < q' & \stackrel{\text{def}}{\Longleftrightarrow} & q = q_{0} \vee
\begin{array}\{{ll}.
\exists (q_{1},\ldots,q_{n},q'_{1},\ldots,q'_{p}), \\
q_{0} \mapsto q_{1} \mapsto \cdots \mapsto q_{n} \mapsto q \mapsto q'_{1}
\mapsto \cdots \mapsto q'_{p} \mapsto q'
\end{array}
\end{eqnarray*}
and $q_{0},q_{1},\ldots,q_{n},q,q'_{1},\ldots,q'_{p},q'$ are
pairwise-different. The notation `$q \le q'$' stands for `$q < q' \vee q =
q'$'. If we consider the $\mathcal{A}_{0}$ automaton pictured at
\ffig~\ref{hstr-automaton}, $q_{0} < q_{1} < q'_{1} < q_{2} < q_{3} < q'_{3} <
q_{4} < q_{5} < q_{6} < q_{8}$ and $q'_{3} < q_{7} < q_{5}$. Obviously, our
`$<$' relation is a partial order.

\begin{remark} \label{already}
In fact, we build a binary relation step by step by exploring all the possible
paths from the initial state, until we reach a state previously explored within
the same chain, and our `\texttt{<}' function is the transitive closure of this
relation. As a consequence, the transitions which do not satisfy this property
are those going back to a state already explored.
\end{remark}

\section{Our Method's Functions} \label{implementation}

\subsection{Our Markers}

Our main idea\Dash already expressed in \cite{h2015a}\Dash is quite comparable
to the \emph{modus operandi} of a model-checker when it checks the successive
states of an automaton in the sense that we mark all the successive states of
a multiple reconfiguration path's automata. The possible values of such a mark
are:
\begin{description}
 \item[\textnormal{\texttt{unchecked}}] the initial mark for the steps not yet
explored within a reconfiguration path;
 \item[\textnormal{\texttt{again}}] if a universal property (for all the
members of a suffix path) is being checked, it must be checked again at this
step if it is explored again;
 \item[\textnormal{\texttt{checked}}] the property has already been checked,
and no additional check is needed if this step is explored again.
\end{description}
However, there is a significant difference between \cite{h2015a} and the
present work: in \cite{h2015a}, one marker was used for a state. This is
impossible here since we have to explore several possible transitions from a
same state. Let us consider the multiple reconfiguration path \textsf{((e
\makebar\ $\mathit{op}_{0}$) $\mathit{op}_{1}$)+}\Dash where
$\mathsf{e},\mathit{op}_{0},\mathit{op}_{1} \in \mathcal{R}_{\mathit{run}}$
with $\mathit{op}_{0} \neq \mathsf{e}$, $\mathit{op}_{1} \neq \mathsf{e}$\Dash
and a property \textbf{after} \textsf{e} \textbf{always} $\mathit{cp}$. When
this regular expression is resumed, there are two cases: either the \textsf{e}
event has been recognised, in which case we have to check the $\mathit{cp}$
property on all the successive states and cycling is detected after the new
application of the \textsf{e} operation, or $\mathit{op}_{0}$ and
$\mathit{op}_{1}$ have been performed and we are still waiting for the
\textsf{e} event. We cannot use the same markers for these two cases.

The type of the \texttt{check-after*} function is given in~\S~\ref{types-used}.
In fact, an automaton modelling a multiple reconfiguration path is
pre-processed and its states are marked as \texttt{unchecked}, by means of a
new mark, \texttt{mark-for-after}. Then a recursive function
\texttt{check-after**}\Dash being the same type\Dash is launched, reads and
updates this new mark. The \texttt{check-always*} function behaves the same,
the recursive function which is launched is \texttt{check-always**} and the new
marker is \texttt{mark-for-always}.

The implementation of the functions \texttt{check-after**} and
\texttt{check-always**} is given in \ffig~\ref{hstr-programs}. We use a
high-level functional pseudo-language, except for updating marks, which is done
by means of side effects. A more complete implementation is available at
\cite{h2017y}, including other features of \logo{ftpl}, with similar
programming techniques and similar methods for proving the termination of our
functions and the correctness w.r.t.\ the definitions given in
\cite{dormoy-etc2010,dormoy-etc2011}.

\begin{figure}[t]

\begin{footnotesize}
\begin{ttfamily}
\begin{tabbing}
ch\= eck-after**($e$,$\mathit{check\text{-}f*}$)($q$,$c$) $\longrightarrow$ \\
 \> if mark-for-after($q$) == again then true \\
 \> el\= se \qquad // mark-for-after($q$) == unchecked \\
 \> \> mark-for-after($q$) $\longleftarrow$ again ; result $\longrightarrow$
true ; \\
 \> \> fo\= r $\tau$ in $\mathtt{t(q)}$ do \\
 \> \> \> $c_{0} \longleftarrow \mathtt{l}(\tau)(c)$ ; $q_{0} \longleftarrow
\tau(q)$ ; \\
 \> \> \> result $\longrightarrow$ \= result and \\
 \> \> \> \> if\=\ l($\tau$) == event->ev-op($e$) and
event->termination-s($e$)($c_{0}$,$c$) then
$\mathit{check\text{-}f*}$($q_{0}$,$c_{0}$) \\
 \> \> \> \> \> else
check-after**($e$,$\mathit{check\text{-}f*}$)($q_{0}$,$c_{0}$) \\
 \> \> \> \> end if \\
 \> \> end for ; \\
 \> \> result ; \\
 \> end if \\
end \\[\smallskipamounttt]
check-always**($\mathit{check\text{-}p*}$)($q$,$c$) $\longrightarrow$ \\
 \> $\mathit{check\text{-}p*}(c)\; \wedge$ \= if mark-for-always($q$) ==
checked then true \\
 \> \> el\= se \qquad // mark-for-always($q$) $\in$ \{unchecked,again\} \\
 \> \> \> mark-for-always($q$) $\longleftarrow$ checked ; result
$\longleftarrow$ true ; \\
 \> \> \> fo\= r $\tau$ in $\mathtt{t}(q)$ do \\
 \> \> \> \> $c_{0} \longleftarrow \mathtt{l}(\tau)(c)$ ; $q_{0} \longleftarrow
\tau(q)$ ; result $\longrightarrow$ result and
check-always**($\mathit{check\text{-}p*}$)($q_{0}$,$c_{0}$) \\
 \> \> \> end for ; \\
 \> \> \> result ; \\
 \> \> end if ; \\
end
\end{tabbing}
\end{ttfamily}

\end{footnotesize}

\caption{Checking properties: two implementations.} \label{hstr-programs}
\frule
\end{figure}

\subsection{Implementations' Correctness} \label{proofs}

Concerning the termination of the functions \texttt{check-after**} and
\texttt{check-always**}, the proofs are similar to those given in
\cite{h2015a}. The correctness is also ensured for idempotent reconfiguration
operations, excluding some operations on parameters, but proofs are here more
subtle.

\subsubsection{Termination}

\begin{proposition} \label{termination}
The function \emph{\texttt{check-after**}} terminates.
\end{proposition}

Let $q_{0}$ be the initial state of our automaton, a principal call of the
\texttt{check-after**} function is:
\begin{center}
\texttt{check-after**}($e$,$\mathit{check\text{-}f*}$)($q_{0}$,$c$)
\end{center}
where $e$ is an event, $\mathit{check\text{-}f*}$ a check function being
\texttt{path-check} type, $c$ a component model. Recursive calls of this
function satisfy the invariant $\forall q_{j} : q_{0} \le q_{j} < q_{i},
\text{\texttt{mark-for-after}}(q_{j}) = \mathtt{again}$ when it is applied to
the $q_{i}$ state. The transitions which may be fired from $q_{i}$ are a finite
set, so the `\textbf{for}' loop terminates if for each transition, the process
terminates. Let $q_{k}$ be a state reached from $q_{i}$. If $q_{i} < q_{k}$,
the invariant holds. If $q_{i} \not< q_{k}$, then $q_{k}$ is a state already
explored\footnote{See \remk~\ref{already}.}, that is, the next recursive call
applies to a state whose the value of \texttt{mark-for-after} is
\texttt{again}. Such a call terminates.

\begin{proposition}
The function \emph{\texttt{check-always**}} terminates.
\end{proposition}

This termination proof is similar: since transitions which may be fired from
$q_{i}$ are a finite set, the `\textbf{for}' loop terminates if for each
transition, the process terminates. However, let us notice that a process
launched by the \texttt{check-always**} function may start after the beginning
of a cycle, and the cycle may have to be entered a second time. Globally, two
passes may be needed for an expression such that
\texttt{check-after*}$(e,\text{\texttt{check-always*}}(cp))$, where $e$ is a
reconfiguration operation and $\mathit{cp}$ a formula. Before reaching the end
of a cycle, the invariant is:
\[\forall q_{j} : q_{0} \le q_{j} < q_{i},
\text{\texttt{mark-for-always}}(q_{j}) = \mathtt{checked} \vee
\text{\texttt{mark-for-always}}(q_{j}) = \mathtt{again}\]
when the \texttt{check-always**} function is applied to the $q_{i}$ state.
Roughly speaking, when a cycle is performed, this mark has been set either to
\texttt{again}, in which case the property has to be checked again, or to
\texttt{checked}, in which case our function concludes that the temporal
property is true. If the mark has been set to \texttt{again}, it means that the
checking of the temporal property `\textbf{always} $\mathit{cp}$' had not begun
yet; for example, if we were processing the `\textbf{after}' part of
`\textbf{after} $e$ \textbf{always} $\mathit{cp}$'. If re-entering a cycle is
needed, at a $q'_{0}$ state already explored, the invariant is $\forall q_{j} :
q'_{0} \le q_{j} < q_{i}, \text{\texttt{mark-for-always}}(q_{j}) =
\mathtt{checked}$, $q_{i}$ being the current state. Let $q_{k}$ a state reached
from $q_{i}$. If $q_{i} < q_{k}$, the invariant holds. If $q_{i} = q'_{0}$,
this recursive call of \texttt{check-always**} is performed with the situation:
\[\forall q_{j} : q'_{0} \le q_{j} \not< q'_{0},
\text{\texttt{mark-for-always}}(q_{j}) = \mathtt{checked}\]
that is, the \texttt{check-always} function terminates at this next call.

\subsubsection{Restrictions on Formulas} \label{cp-flat-modus-operandi}

Let us recall that in \cite{h2015a}, we were able to deal with finite paths and
cycles without continuation, that is, the `\texttt{+}' construct of regular
expressions was used only at a final position. In other words, there were no
alternatives. In this previous work, we also mentioned that our \emph{modus
operandi} is suitable if the cycle of reconfiguration operations is
\emph{idempotent}. Since the composition of two commutative idempotent
functions is idempotent, too, some pairs of reconfiguration operations can be
commuted, some consists of operations which neutralised each other, and
globally, most cycles used are globally idempotent. Concerning our primitive
reconfigurations, most of them are idempotent, e.g., a component's addition or
removal, as well as a binding's addition or removal. Assigning a constant
value to a parameter is idempotent, but general changes are not, e.g.,
incrementing or decrementing a parameter.\smallskip

Of course, this limitation still holds for our revised algorithms. Another
limitation exists for alternative with a common continuation. As a simple
counter-example, let us consider the multiple reconfiguration path
\texttt{($\mathit{op}_{0}$ \makebar\ $\mathit{op}_{1}$) $\mathit{op}_{2}$}. If
we process the formula \textsf{always} $\mathit{cp}$\Dash $\mathit{cp} \in
\mathit{CP}$\Dash our algorithm checks the $\mathit{cp}$ formula at the initial
state, then at the result of $\mathit{op}_{0}$, then at the result of
$\mathit{op}_{2}$ after $\mathit{op}_{0}$. The result of $\mathit{op}_{1}$
applied to the initial state is checked, and the process stops because of the
mark put at the common state after $\mathit{op}_{0}$ and $\mathit{op}_{1}$. Now
let $\mathit{cp}$ be $\mathit{cp}_{0} \vee \mathit{cp}_{1}$\Dash where
$\mathit{cp}_{0},\mathit{cp}_{1} \in \mathit{CP}$\Dash and let us assume that
$\mathit{cp}_{0} \wedge \neg\mathit{cp}_{1}$ (resp.\ $\neg\mathit{cp}_{0}
\wedge \mathit{cp}_{1}$) holds on the result of $\mathit{op}_{0}$ (resp.\
$\mathit{op}_{1}$). If $\mathit{cp}_{1}$ is always false after applying
$\mathit{op}_{2}$\Dash e.g., $\mathit{cp}_{1}$ may be related to a binding
removed by $\mathit{op}_{2}$\Dash, our method results in an erroneous answer
along the path $\mathit{op}_{1}$ $\mathit{op}_{2}$, even it is right for the
path $\mathit{op}_{0}$ $\mathit{op}_{2}$.

Solutions exist. We could restrict alternatives of regular expressions by
allowing them only at the top level. The counter-example above would be
rewritten as \texttt{($\mathit{op}_{0}$ $\mathit{op}_{2}$ \makebar\
$\mathit{op}_{1}$ $\mathit{op}_{2}$)}, the result of $\mathit{op}_{2}$\Dash as
a component model\Dash would be checked twice, one time after applying
$\mathit{op}_{0}$, the second after applying $\mathit{op}_{1}$. Adopting such a
rule would complicate the processing of a multiple reconfiguration path such as
\texttt{($\mathit{op}_{0}$ \makebar\ $\mathit{op}_{1}$)+}. Another drawback is
that a multiple reconfiguration path may contain alternatives for the
corresponding automaton even if the `\makebar' operator is not used explicitly.
As an example, let us consider the multiple reconfiguration path
\texttt{$\mathit{op}_{0}$ $\mathit{op}_{1}$?\ $\mathit{op}_{2}$}. The
alternative syntactically appears if we rewrite it by means of a grammar\Dash
$S$ being the axiom, $S'$ another non-terminal symbol, and $\varepsilon$ the
empty word\Dash:
\begin{center}
\texttt{S $\longrightarrow$ $\mathit{op}_{0}$ S' $\mathit{op}_{2}$ \qquad
\qquad S' $\longrightarrow$ $\mathit{op}_{1}$ \makebar\ $\varepsilon$}
\end{center}
and an analogous counter-example, based on a logical disjunction, can be found
for such a case. This drawback does not appear if a \emph{non-empty} cycle is
possibly followed by a continuation, that is, in a multiple reconfiguration
path like \texttt{$\mathit{op}_{0}$+ $\mathit{op}_{1}$}. If we rewrite this
example by means of a grammar:
\begin{center}
\texttt{S $\longrightarrow$ $\mathit{op}_{0}$ S' \qquad \qquad S'
$\longrightarrow$ $\mathit{op}_{0}$ S' \makebar\ $\mathit{op}_{1}$}
\end{center}
we will see that no common part follows the alternative. This is different if
the cycle can be empty. As an example, the multiple reconfiguration path
\texttt{$\mathit{op}_{0}$ $\mathit{op}_{1}$* $\mathit{op}_{2}$} can be
rewritten using the following grammar:
\begin{center}
\texttt{S $\longrightarrow$ $\mathit{op}_{0}$ S' $\mathit{op}_{2}$ \qquad
\qquad S' $\longrightarrow$ $\mathit{op}_{1}$ S' \makebar\ $\varepsilon$}
\end{center}
and a common part follows the alternative.\smallskip

From our point of view, the best solution is to restrict formulas to the strict
subset $\mathit{CP}^{\flat}$ defined in \S~\ref{cp-flat}. In other words, the
`$\vee$' connector must not be used, the `$\forall$' quantifier\Dash related to
that connector\Dash and the `$\neg$' operator must not, either.

\subsubsection{Correctness for Restricted Formulas}

Adopting these additional conventions, proving the correctness of our function
\texttt{check-always*}\Dash other functions' correctness is analogouss\Dash is
tedious but not really difficult. We have to examine all the basic cases of
formulas $\mathit{cp} \in \mathit{CP}^{\flat}$)\Dash e.g., the set membership
of a binding\Dash and idempotent reconfiguration operations
$\mathit{op}_{0},\mathit{op}_{1},\mathit{op}_{2}$ to show the following
proposition.

\begin{proposition}
Starting from the same state and the same component model, if the formula
\textnormal{\textsf{always $\mathit{cp}$}}\Dash where $\mathit{cp} \in
\mathit{CP}^{\flat}$\Dash holds on the two paths $\mathit{op}_{0}$
$\mathit{op}_{2}$ and $\mathit{op}_{1}$\Dash that is, before and after applying
$\mathit{op}_{1}$\Dash it also holds on the multiple reconfiguration path
\emph{\texttt{($\mathit{op}_{0}$ \makebar\ $\mathit{op}_{1}$)
$\mathit{op}_{2}$}}.
\end{proposition}

By induction, it is easy to prove such a property about longer paths. It is
also easy to prove that if this property holds for the two formulas
$\mathit{cp}_{0}$ and $\mathit{cp}_{1}$, it also holds for the formula
$\mathit{cp}_{0} \wedge \mathit{cp}_{1}$. An analogous proof exists for the 
`$\forall$' quantifier. By induction on the number of members of a multiple
reconfiguration path, we can prove this proposition by considering a grammar
associated with this path, as we sketch in \S~\ref{cp-flat-modus-operandi}. As
a consequence, if a same state is reached along several paths, the property
holds and our function \texttt{check-always**} is correct. Studying the
correctness of the function \texttt{check-after**} is easier, because the
possible futures of each path of an alternative are explored independently.

\section{Discussion and Future Work} \label{discussion}

Within the framework sketched at~\S~\ref{modus-operandi}, the new versions of
our programs have been implemented using the \pgname{Java} programming
language and can be found in \cite{h2017y}. The descriptions of this paper
allow us to be more related to a theoretical model, and to emphasise that our
method is close to algorithms based on marking tehniques and used in
model-checking, e.g., \cite{clarke-etc1986,clarke-etc1994,queille-sifakis1982}.

As mentioned in the introduction, our method takes place at design-time. We do
not deal with a language to describe reconfiguration operations and constraints
on these operations as an extension of an \logo{adl}, as in
\cite{sanchez-etc2015}, we are mainly interested in developing \emph{effective}
methods for verifying properties. In \cite{h2015a} we were able to deal with a
particular case of infinite paths, based on the fact that often the same
sequences are repeated: a component may be stopped in some circumstances,
restarted in some circumstances, and so on. However, it is true that this
situation was restrictive and the initial motivation of the present work was to
introduce alternatives within our paths. Such construct would be irrelevant
within methods working at run-time
\cite{kouchnarenko-weber2013,kouchnarenko-weber2014}, since they observe a
process in progress, the history of reconfiguration operations being known. At
design-time, it may be interesting to plan several possible behaviours, what is
new in comparison with \cite{h2015a}. In the present work, we choose to focus
on some efficiency for our algorithms, since common parts are explored once and
cycles are explored two times at most, that is, our algorithms are linear with
respect to the automaton's state number. In other words, we are able to explore
several possible behaviours quite efficiently, but the price to pay is a
restriction of the formulas processed. However, if we look at the examples
given within
\cite{dormoy-etc2010,dormoy-etc2011,dormoy-etc2012,kouchnarenko-weber2013}, we
can think that our restriction is not too cumbersome in practice.

As mentioned above, other solutions exist, but we wanted our extension to be
close to our original \emph{modus operandi}. If we consider a `simple'
reconfiguration path, that is, only one transition starts from each state of
the corresponding automaton, we get exactly the programs given in
\cite{h2015a}. Yet another work may consider only alternatives without
syntactic common continuation\Dash possibly by applying some transformation
rules\Dash or our algorithms could be changed in order to explore more states
in such a case, but this second solution might lead to some combinatorial
explosion. Another solution could be based on \emph{branching-time} logic for
reconfiguration alternatives, whereas the present work is based on linear-time
logic, as in
\cite{dormoy-etc2010,dormoy-etc2011,dormoy-etc2012,kouchnarenko-weber2013}.
Other ideas could be based on a connection with the Model Driven Engineering
technical space \cite{bezivin2006}, who would provide more expressive power.
Likewise, we could plan a bridge between our approach and others, closer to a
semantic level: for example, \cite{krause-etc2011} models reconfiguration
operations by means of graph rewriting and uses formal verification techniques
along graphs to check properties related to reconfigurations.

On another point, we are interested in this work in reconfigurations, but not
in \emph{reasons} for these reconfigurations\footnote{This is the same in
\cite{h2015a}.}, most often expressed by \emph{reconfiguration policies}
\cite{chauvel2008a}. In parallel, we are working on an extension of
\cite{h2015a} taking such policies into account \cite{h2017yy}. In the future,
we plan to integrate reconfiguration policies into our approach based on
mutiple reconfiguration paths.

\section{Conclusion}

In comparison with methods at run-time, ours may appear as too static, unable
to cope with unexpected situations. Our plan is to investigate as far as
possible properties that can be checked at design-time, in order for a
reconfigurable system to be deployed as safely as possible. Our work can be
used for simulations, it may help conceptors design policies involving
reconfigurations with good properties. Our tool is not ready for testing
policies, but can be used for testing \emph{possible results} of policies. We
see that such an approach does not aim to replace works applied at run-time,
but to complement them. About examples such as an \logo{http} server, we
succeeded in proving properties. In other words, we think that our method can
provide some significant help at design-time.

\section*{Acknowledgements}

I am grateful to Olga Kouchnarenko and Arnaud Lanoix, who kindly permitted me
to use \ffigs~\ref{hstr-http-server} \&~\ref{hstr-http-path}. Many thanks to
the anonymous referees, who pointed out some omissions and suggested me
constructive improvement.

\bibliographystyle{eptcs}
\bibliography{C-et-al,hufflen-2010-,more-xml,new,nfp,other-languages,plus}

\end{document}